# The economic value of empowering older patients transitioning from hospital to home: Evidence from the 'Your Care Needs You' intervention

Alfredo Palacios[1*], Simon Walker[1], Beth Woods[1], Catherine Hewitt[2], Alison Cracknell[3], Jenni Murray[4], Rebecca Lawton[4,5], Gerry Richardson[1]

[1] Centre for Health Economics (CHE), University of York, York, UK

[2] York Trials Unit, University of York, York, UK

[3] Leeds Centre for Older People's Medicine, Leeds Teaching Hospitals NHS Trust, Leeds, UK

[4] Yorkshire Quality and Safety Research Group, Bradford Institute for Health Research, Bradford, UK

[5] School of Psychology, University of Leeds, Leeds, UK

**Version information:**

- First version: 24th October 2024
- Current version: 24th November 2024
- To access the most recent version of this manuscript, please click here

**Important Note:**

This is a work in progress. Please do not cite or circulate without the authors' permission.

[*] Corresponding author. Email: alfredo.palacios@york.ac.uk. We acknowledge the comments and suggestions of Robbie Foy, Stephen Brealey, Laura Mandefield, Ruth Baxter, Kalpita Baird, Jane O'Hara, Laura Sheard, Ed Breckin, and Lubena Mirza on the design, implementation, and analysis of this study. We also acknowledge the valuable feedback provided by anonymous reviewers of the NIHR and the participants of the HESG Summer Meeting at the University of Warwick 2024.

# Abstract

Hospital-to-home transitions are a critical component of effective healthcare delivery, especially for patients aged 75 and older. This study evaluates the cost-effectiveness of the 'Your Care Needs You' (YCNY) intervention, a patient-centred approach designed to empower older adults during discharge, compared to standard care. The analysis adopts the perspective of the National Health Service (NHS) and Personal Social Services. Data were drawn from a cluster randomised controlled trial (cRCT) conducted within the UK NHS over a 90-day post-discharge follow-up period. Adjusted differences in costs and Quality-Adjusted Life Years (QALYs) were estimated using Multilevel Mixed-Effects Generalised Linear Models (MME-GLM) to account for the hierarchical structure of the trial design. Alternatively, Seemingly Unrelated Regression (SUR) models were employed to address potential correlations between costs and QALYs. Scenario analyses and probabilistic sensitivity analyses were conducted to assess the robustness of the results. The YCNY intervention reduced costs by £269 and achieved a QALY gain of 0.0057, resulting in a net health benefit (NHB) of 0.0246 at a £15,000/QALY threshold. It demonstrated an 89% probability of cost-effectiveness compared to standard care within the trial's time horizon. Findings remained robust across alternative scenarios and sensitivity analyses. These results suggest that YCNY is a promising and potentially cost-effective strategy for improving hospital-to-home transitions for older adults. By enhancing outcomes and reducing costs, the study supports integrating patient-centred interventions like YCNY into routine NHS practice, with the potential to improve both efficiency and quality of healthcare delivery.

1. **Introduction**

The transition from hospital to home is a vulnerable period for elderly patients, marked by a high risk of emergency readmissions that can lead to poor health outcomes (Mor and Besdine, 2011; Naylor et al., 1999). In the United Kingdom National Health Service (NHS), 19.6% of patients aged 75 and over are readmitted to hospitals in an emergency within 30 days of their initial discharge (NHS, 2022). Notably, a high proportion of these readmissions, approximately 30%, are considered preventable with more effective transitional care and support systems (Auerbach et al., 2016; Blunt et al., 2015; van Walraven et al., 2011). This significant rate of emergency readmission not only imposes a substantial financial burden on the healthcare system but also compromises the quality of life for this patient group, underscoring the urgent need for effective transitional care interventions.

The 'Your Care Needs You' (YCNY) intervention is designed to empower patients by providing them with crucial knowledge and tools essential for managing their care following hospital discharge. The YCNY intervention comprises three structured components directed at patient engagement: (i) a booklet that encourages patient involvement in their healthcare and provides guidance across four key areas—health monitoring, medication management, daily activities, and steps for escalation of care when needed, (ii) an informational film that reinforces the importance of the YCNY intervention and addresses patients' health beliefs and emotional concerns, and (iii) a comprehensive care summary given to patients at discharge, which includes practical information and resources for social support. Additionally, hospital staff are encouraged to facilitate patient engagement with these resources and to provide the necessary social and emotional support to enhance the effectiveness of the YCNY intervention (Baxter et al., 2020).

The YCNY intervention was implemented and assessed through a cluster randomised controlled trial (cRCT) across forty NHS wards in England (ISRCTN17062524). Wards were allocated to either the YCNY intervention group or the care-as-usual control group, with randomisation stratified by ward specialty and patient characteristics. The primary clinical outcome was the rate of unplanned 30-day readmissions, with secondary outcomes including readmissions at 60 and 90 days, time to readmission, and readmission duration. While no statistically significant difference was observed for unplanned 30-day readmissions (OR 0.93; 95% CI, 0.78–1.10; p=0.372), the intervention group demonstrated consistently lower readmission rates across all timepoints. At 90 days, a significant 13% reduction in the total number of readmissions was observed in the YCNY group (IRR 0.87; 95% CI, 0.76–0.99; p=0.039). These results suggest a benefit of the YCNY intervention in improving post-discharge care and reducing readmissions. Although these findings are promising in terms of patient outcomes, the economic implications of adopting such transitional care models remain unexplored.

A systematic review of economic evaluations focusing on interventions aimed at preventing hospital readmissions identified 50 relevant studies (Nuckols et al., 2017). Among these, 14 specifically addressed interventions targeting adults aged 65 and older. Seven of these studies concentrated on individuals with a specific health condition, such as heart failure, while the remaining seven encompassed a general population of older adults. The findings from these seven economic studies focused on the general elderly population indicated that the interventions were generally cost-saving (Gillespie et al., 2009; Graham et al., 2015; Graves et al., 2009; Jayadevappa et al., 2006; Naylor et al., 1994, 1999; Saleh et al., 2012). However, it is important to note that six out of these seven studies were cost-comparison analyses rather than full economic evaluations, thereby limiting the comprehensiveness of their economic assessments. This highlights a gap in comprehensive economic evaluations within this critical area of healthcare, underscoring the need for more detailed and methodologically rigorous studies to better inform policy and practice.

The aim of this study is to conduct a within-trial economic evaluation of theYCNY intervention from the perspective of the NHS and Personal Social Services (PSS). In line with the trial's design, this evaluation covered the 90-day period post-discharge to provide an analysis of both YCNY intervention and healthcare costs, alongside Quality-Adjusted Life Years (QALYs). This evaluation is essential for the NHS as it seeks to ascertain the cost-effectiveness of the YCNY intervention, thereby determining its 'value for money' and impact on population health. The findings will be crucial in guiding resource allocation to maximise population health and facilitate the integration of patient-centred transitional care models within the healthcare system.

2. Methods

*Study design*

The economic analysis utilised individual patient-level data from the YCNY trial to assess the cost-effectiveness of the YCNY intervention versus usual care. These calculations include YCNY intervention costs, healthcare costs and health outcomes. A cost-effectiveness analysis was conducted which calculated the incremental cost-effectiveness ratio (ICER) based on the adjusted mean differences in costs and QALYs between the YCNY intervention and control groups. Additionally, the net health benefit (NHB) and net monetary benefit (NMB) were calculated using a cost-effectiveness threshold of £15,000 per QALY gained (Martin et al., 2023). A health economics analysis plan (HEAP) was developed prior to conducting the economic evaluation to outline the specific methods and procedures. The HEAP is available upon request.

The analysis was conducted from the perspective of the NHS and Personal Social Services (PSS), aligning with UK guidelines for economic evaluation (NICE, 2022). In line with the clinical trial, the within-trial economic evaluation focuses on a 90-day post-randomisation period. Given the trial's short duration, costs and benefits were not discounted. Cost-effectiveness thresholds of £15,000 (Martin et al., 2023), and £20,000 and £30,000 per QALY, as recommended by NICE (2022), were used to evaluate the cost-effectiveness of the YCNY intervention. The analysis adhered to established guidelines for reporting within-trial economic evaluation studies (Gomes et al., 2012a; Husereau et al., 2022; Ramsey et al., 2015) to ensure comprehensive and transparent reporting. CHEERS 2022 Checklist is presented in Table S1 of the Supplementary material.

*Participants*

The inclusion and exclusion criteria for the YCNY intervention evaluation are described in detail elsewhere. Briefly, participants were eligible if they were aged 75 or older, expected to be discharged to their own or a relative's home, had stayed at least one night on a participating ward, and were able to provide informed consent. Participants were excluded if they required an interpreter, lived outside the study area, or were expected to transfer to another acute hospital or community rehabilitation unit, among other reasons.

Patients who died within 30 days without readmission were excluded from the primary clinical analysis. However, these patients were included in the economic evaluation to mitigate potential bias, contributing to both cost and QALY calculations. A summary of the sample sizes across datasets is available in Table S2 in the Supplementary Material.

*Data*

The YCNY trial utilises two primary data sources: 1) routine data, which includes variables such as unplanned hospital readmissions at 30, 60, and 90 days, length of hospital stays, ward characteristics, and other relevant variables; and 2) Case Report Form (CRF) data, which contains patient characteristics and EQ-5D measurements and some resource use utilisation (some hospital services, and primary care and social care) for a subsample of individuals from the routine data. These EQ-5D and resource use measurements were taken at baseline (collected in hospital during the initial admission and following patient recruitment), and at 10, 30, and 90 days post-discharge.

In order to consolidate the unplanned hospital readmissions, the EQ-5D and resource use measurements, all essential for the economic evaluation, we combined both datasets (Routine and CRF datasets) creating a Merged Dataset. Therefore, only patients contributing to both datasets (Routine and CRF datasets) were included in the analysis. The analysis of this Merged dataset followed the "intention-to-treat" (ITT) principle.

*Intervention and resource use measures and valuation*

Resource use data collection involved the following categories/resources post initial discharge: 1) hospital services: unplanned readmissions at 30, 60, and 90 days and the duration of inpatient stays, outpatient clinic, day case, and accident and emergency (A&E); 2) primary care services: GP at surgery, home and telephone, nurse at surgery, home and telephone, and therapist; and 3) social services: home care and social worker. While unplanned readmissions at 30, 60, and 90 days and the duration of inpatient stays comes from a routine dataset, the information for rest of the healthcare and social services comes from the CRF to patients and caregivers.

Regarding the YCNY intervention cost, we considered the staff profile and time to be trained and to provide the YCNY intervention. This information was captured in a separate administrative documentation based on expert opinion. The staff profile involved mainly nurses/managers, and the main activities for delivering the YCNY intervention were to discuss with patients about care during admission and discharge, assisting with activities of daily living and providing instructions/education for patient and/or caregiver. For further details about the delivery of the YCNY intervention see Table S3 in the Supplementary material.

All resource utilisation was monetized and expressed in 2022 British pounds, using published UK unit cost estimates (Jones et al., 2023; NHS, 2023). The cost of implementing the YCNY intervention was calculated based on staff salary data (Jones et al., 2023).

*Outcome measures and valuation*

The primary economic outcome is Quality-Adjusted Life Years (QALYs), derived from health-related quality of life scores obtained through the EQ-5D-5L instrument. These scores were mapped to the EQ-5D-3L value set using the Hernandez-Alava et al. (2022) mapping algorithm, following UK guidance. Outcome measurements were collected at baseline (during the initial hospital admission and post-recruitment) and at 10, 30, and 90 days post-discharge (via postal questionnaires). QALYs were calculated using the area under the curve approach.

*Missing data*

The nature and extent of the missing data were assessed to determine the most appropriate imputation method. For missing data at baseline, within-cluster mean imputation was used independently of the YCNY intervention allocation (Faria et al., 2014; Taljaard et al., 2008).

In our base case, we assumed that data was missing at random and multiple imputation methods through chained equations (MICE) by arm were employed (Faria et al., 2014; White et al., 2011). The number of imputed datasets was defined according to the percentage of incomplete cases (White et al., 2011), and the analysis was conducted using the 'mi impute' package in Stata. In the imputation model the same variables as in the main analysis were included and the hierarchical structure of the data incorporated (Díaz-Ordaz et al., 2013; Gomes et al., 2013; Graham, 2009).

For the specific case of EQ-5D variables, the decision about imputing the domains or the EQ-5D index was based on the sample size and the proportion of missing data (Rombach et al., 2018; Simons et al., 2015). Given the potential situation of missing values for the resource use information coming from CRFs, we conducted multiple imputation for resources with missing values lower 60%, and these resources were part of the main cost-effectiveness analysis. For the rest of resources, that is, those with a percentage of missing values higher than 60%, a resource use comparison between YCNY intervention and usual care was presented considering a complete case analysis.

*Statistical methods*

Mean costs and QALYs for each group were presented alongside their adjusted mean differences, with 95% confidence intervals around the differences in means (Willan et al., 2004). In accordance with the clinical outcomes, the mean differences for costs were controlled by ward specialty, readmission rate at baseline (ward level variable), percentage of patients over 75 years (ward level variable), and sex (patient level variable). The mean differences for QALYs were controlled by EQ-5D at baseline and the same variables as the analysis of mean difference for costs.

For the base case analysis, multilevel mixed-effects generalised linear models (MME-GLM) were estimated for analysing differences in mean costs and outcomes considering the hierarchical structure of the data (wards in hospitals as clusters) (Gomes et al., 2012b). Seemingly unrelated regression (SUR) models with robust standard errors were also considered to account for potential intragroup (cluster) correlations between costs and QALYs. Costs and QALYs were combined to calculate an incremental cost-effectiveness ratio (ICER), as well as the net health benefit (NHB) using the cost-effectiveness threshold of £15,000 per QALY (Martin et al., 2023).

Non-parametric bootstrap was also used to produce the cost-effectiveness plane, representing the uncertainty in incremental cost and effect estimates, and the probability of YCNY intervention being cost-effective at different thresholds. Given that different approaches have been proposed to combine multiple imputation and bootstrap, we followed a possible approach consisting of drawing bootstrap samples from each of the imputed dataset separately and then pooling the estimates (Leurent et al., 2018; Schomaker and Heumann, 2018).

*Alternative scenario analyses*

In the base case, we assumed that the missingness mechanism for HRQoL data followed a Missing at Random (MAR) process. To assess the robustness of this assumption, we conducted scenario analyses exploring potential departures from MAR for HRQoL data, while retaining the MAR assumption for missing cost data. Using the framework proposed by Leurent et al. (2018), we applied a systematic approach to simulate Missing Not at Random (MNAR) scenarios for HRQoL data.

Initially, we imputed the missing HRQoL values under the MAR assumption using multiple imputation (MI). Subsequently, we adjusted the MAR-imputed HRQoL values to reflect plausible MNAR scenarios by proportionally modifying the imputed values within each trial arm by ±5% and ±10%, resulting in seven alternative scenarios. These adjusted datasets were then analyzed using conventional methods for multiple imputed data to quantify the impact of deviations from the MAR assumption.

3. Results

*Participants*

Table 1 presents the baseline characteristics of 468 participants in 35 hospital wards included in the Merged dataset, divided into YCNY intervention (patients=222, wards=16) and control (patients=246, wards=19) groups. Both groups are similar in terms of the individual patient level characteristics, such as age and gender. Some differences are observed in the ward level variables: the YCNY intervention group shows a slightly lower baseline readmission rate and percentage of patients over 75 years old compared to the control group. For an analysis comparing the baseline patient and ward characteristics between the Routine dataset and the Merged dataset, please see Table S4 in the Supplementary material. For a description of the missing data by intervention arms, see Table S5 in the Supplementary material.

**Table 1.** Baseline characteristics.

| Variable | YCNY Intervention | Control | Total | p-value |
|---|---|---|---|---|
| | Patients=222 Wards=16 | Patients=246 Wards=19 | Patients=468 Wards=35 | |
| Age | 82.8 (5.4) | 83.4 (5.5) | 83.1 (5.4) | 0.24 |
| Dummy sex, 1=Male, % | 48.2 (50.1) | 39.4 (49.0) | 43.6 (49.6) | 0.056 |
| EQ-5D at baseline | 0.50 (0.31) | 0.48 (0.33) | 0.49 (0.32) | 0.44 |
| Ward: baseline readmission rate, % | 18.4 (6.6) | 19.7 (5.6) | 19.1 (6.1) | 0.028 |
| Ward: patients over 75 yo, % | 69.5 (27.6) | 74.3 (19.3) | 72.0 (23.7) | 0.028 |
| Ward: Dummy specialty, 1=Eldery & interm. care, % | 53.2 (50.0) | 55.7 (49.8) | 54.5 (49.9) | 0.58 |

*Healthcare resource use and costs*

Table 2 presents the number and percentage of unplanned readmissions within the trial follow-up period (90 days) for both the YCY intervention and control groups. In the YCNY intervention group 73% of patients had zero readmissions, compared to 67% in the control group. The percentage of participants with one readmission was higher in the control group (23%) than in the YCNY intervention group (18%). Both groups were similar in the percentage of participants with two readmissions, at approximately 7%. The control group had a higher percentage of participants with three or more compared to the YCNY intervention group.

**Table 2.** Hospital readmissions within 90 days by arm. Data are presented as N (%)

| Readmissions | YCNY intervention | Control | Total |
|---|---|---|---|
| 0 | 162 (72.97) | 166 (67.48) | 328 (70.09) |
| 1 | 41 (18.47) | 56 (22.76) | 97 (20.73) |
| 2 | 16 (7.21) | 18 (7.32) | 34 (7.26) |
| 3 | 2 (0.9) | 5 (2.03) | 7 (1.5) |
| 4 | 1 (0.45) | 0 (0) | 1 (0.21) |
| 5 | 0 (0) | 1 (0.41) | 1 (0.21) |
| Mean (SD) | 0.3739 (0.6990) | 0.4553 (0.7746) | 0.4167 (0.7400) |

An unadjusted comparative analysis of the YCNY intervention costs, healthcare costs, and total costs between the YCNY intervention and control groups suggest that the mean YCNY intervention cost was £94.32, while the readmission hospitalisation costs averaged £1,502 in the YCNY intervention group and £1,812 in the control group, resulting in reduction of £310. Overall, the total costs for the YCNY intervention group were £1,597, compared to £1,812 for the control group, reflecting a decrease of £216 (for further details see Table S6 in the Supplementary material). For a comparative analysis of other hospital services, primary care and social care services between arms based on a complete case analysis, see Table S7 in the Supplementary material.

*Health outcomes*

Table 3 presents the unadjusted mean EQ-5D-3L scores across four time points: baseline, 10 days, 30 days, and 90 days, and QALYs comparing the YCNY intervention and control arms based on imputed values. For a description of the missing values pattern EQ-5D-3L scores see Table S8, Figure S1 and Table S9 in the Supplementary material.

At baseline, the mean EQ-5D-3L scores were similar between the groups, with the YCNY intervention arm slightly higher. Over time, the YCNY intervention group consistently showed higher mean EQ-5D scores compared to the control group.

**Table 3.** Unadjusted mean for EQ-5D-3L and QALYs variables between arms. Results based on imputed data.

| Variable | YCNY Intervention (N=222) | Control (N=246) |
|---|---|---|
| EQ-5D at Baseline | 0.5010 | 0.4782 |
| EQ-5D at 10 days | 0.4788 | 0.4498 |
| EQ-5D at 30 days | 0.4747 | 0.4384 |
| EQ-5D at 90 days | 0.4945 | 0.4156 |
| QALYs | 0.1191 | 0.1072 |

*Cost-effectiveness*

Table 4 presents a summary of the main cost-effectiveness results. For our base case (MME-GLM), the adjusted differences in total costs and total QALYs were -£268.78 [95% CI: -889.67 - 352.11] and 0.0057 [95% CI: -0.0108 - 0.0223], respectively. In the alternative specification (SUR model), the adjusted differences in total costs and total QALYs were -£233.75 and 0.0071, respectively (the unadjusted mean total costs and total QALYs by arm are presented in Table S10 in the Supplementary material). Given that the YCNY intervention is associated with lower costs and a slight QALY increment, the YCNY intervention could be considered "Dominant" relative to usual care.

**Table 4**. Cost-effectiveness results. Monetary values expressed in British pounds of 2022. Results based on imputed data.

|  |  | Base case: MME-GLM | Alternative case: SUR model |
|---|---|---|---|
| Total costs | Adj. Mean Control* | 1,574.29 | 1,576.43 |
|  | Adj. Mean YCNY intervention* | 1,305.51 | 1,342.68 |
|  | Adj. Mean Difference* | -268.78 | -233.75 |
|  | SE Difference | 316.78 | 333.57 |
|  | 95% CI Difference | -889.67 - 352.11 | -887.54 - 420.04 |
| Total QALYs | Adj. Mean Control** | 0.0835 | 0.0841 |
|  | Adj. Mean YCNY intervention** | 0.0892 | 0.0912 |
|  | Adj. Mean Difference** | 0.0057 | 0.0071 |
|  | SE Difference | 0.0084 | 0.0077 |
|  | 95% CI Difference | -0.0108 - 0.0223 | -0.0080 - 0.0222 |
| ICER |  | YCNY Dominant | YCNY Dominant |

*Notes*: *Results adjusted by ward specialty, readmission rate at baseline (ward level variable), percentage of patients over 75 years (ward level variable), and sex (patient level variable). **Results adjusted by EQ-5D at baseline ward specialty, readmission rate at baseline (ward level variable), percentage of patients over 75 years (ward level variable), and sex (patient level variable). MME-GLM: Multilevel mixed-effects generalised linear models, SUR: Seemingly unrelated regression, QALY: quality-adjusted life year, ICER: incremental cost-effectiveness ratio.

*Scenario analysis*

Table 5 presents the cost-effectiveness of the YCNY intervention under various assumptions regarding missing not at random (MNAR) quality-of-life data, while Figure 1 shows the corresponding cost-effectiveness planes (the corresponding cost-effectiveness acceptability curves are presented in Figure S2 in Supplementary material). The table compares different scenarios by rescaling the MNAR parameters for both the control and YCNY groups. In the base case scenario (Scenario 1), which assumes missing at random (MAR), the YCNY intervention is dominant with an incremental cost of -£268.78 and an incremental QALY gain of 0.0057, resulting in an 89% probability of being cost-effective. Other scenarios adjust the rescaling parameters, for example, reducing imputed quality-of-life values by 5% or 10%. Across all scenarios (Scenarios 2-7), the YCNY intervention consistently remains dominant, with incremental QALY gains between 0.0027 and 0.0058. The probability of cost-effectiveness at a threshold of £15,000 per QALY varies from 87% to 90%, underscoring the robustness of the YCNY intervention's cost-effectiveness under different assumptions of missing data. All of these results are based on imputed data.

Table 5. Cost-effectiveness of YCNY intervention under different MNAR assumptions for missing quality-of-life data. Results based on imputed data.

| Scenario number | MNAR rescaling parameters[a] | | Incremental cost[b] (£) [95% CI] | Incremental QALYs [95% CI] | NHB [95% CI] | Prob. cost-effective[c] (%) |
|---|---|---|---|---|---|---|
| | C control | C YCNY | | | | |
| 1 (Basecase, MAR) | 1 | 1 | -268.78 [-889.67 - 352.11] | 0.0057 [-0.0108 - 0.0223] | 0.0246 [-0.0190 - 0.0682] | 89 |
| 2 | 1 | 0.95 | -268.78 [-889.67 - 352.11] | 0.0027 [-0.0135 - 0.0189] | 0.0231 [-0.0205 - 0.0667] | 87 |
| 3 | 0.95 | 1 | -268.78 [-889.67 - 352.11] | 0.0057 [-0.0103 - 0.0216] | 0.0261 [-0.0174 - 0.0670] | 90 |
| 4 | 0.95 | 0.95 | -268.78 [-889.67 - 352.11] | 0.0043 [-0.0117 - 0.0202] | 0.0246 [-0.0189 - 0.0681] | 89 |
| 5 | 0.95 | 0.90 | -268.78 [-889.67 - 352.11] | 0.0029 [-0.0130 - 0.0187] | 0.0231 [-0.0204 - 0.0666] | 87 |
| 6 | 0.90 | 0.95 | -268.78 [-889.67 - 352.11] | 0.0058 [-0.0098 - 0.0214] | 0.0261 [-0.0174 - 0.0656] | 90 |
| 7 | 0.90 | 0.90 | -268.78 [-889.67 - 352.11] | 0.0044 [-0.0111 - 0.0200] | 0.0246 [-0.0189 - 0.0680] | 89 |

*Notes*: All results are based on imputed data and comparing the YCNY intervention arm to the control arm (n = 468). NHB: net health benefit, MAR: missing at random, MNAR: missing not at random, QALY: quality-adjusted life year, YCNY: Your Care Needs You. CI: confidence interval. [a]How missing quality-of-life data are assumed to differ from the MAR-imputed values, for example, c control = 0.9 means that all imputed quality-of-life values in the control arm have been reduced by 10%. [b]Missing costs assumed to be MAR in all scenarios. [c]At a cost-effectiveness threshold of £15,000/QALY.

**Figure 1**. Cost-effectiveness planes of the YCNY intervention under different MNAR assumptions. Results based on imputed data.

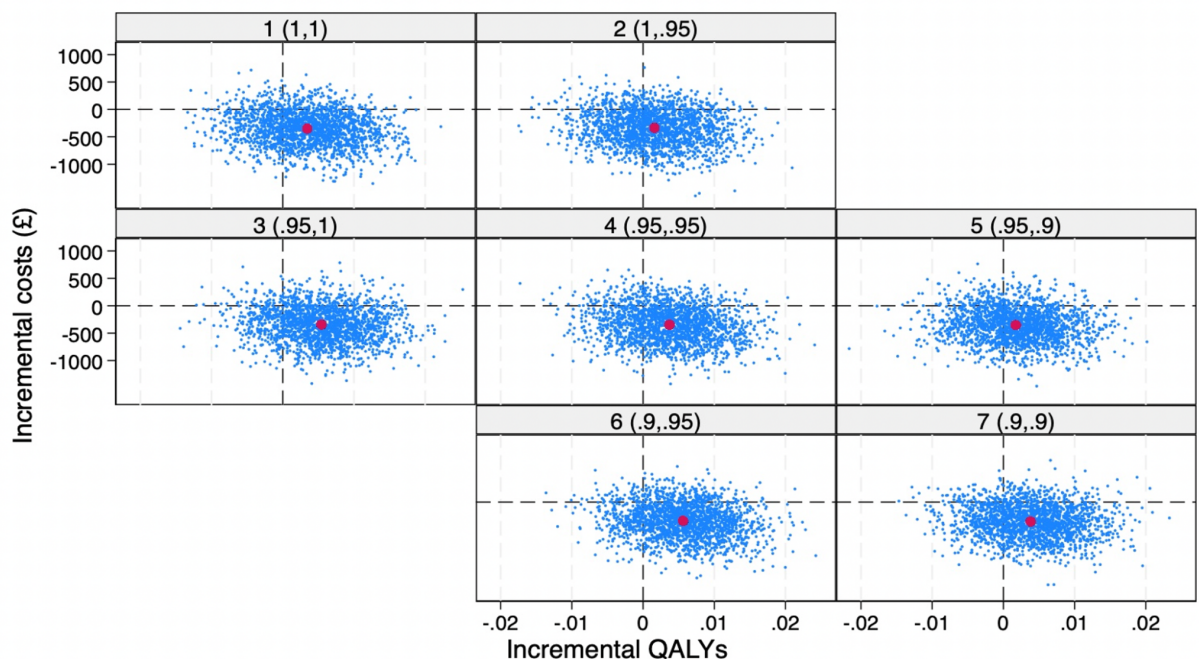

Graphs by scenario

*Notes*: Headings in the top of each plane indicate the scenario number and the MNAR rescaling parameters (c control, c YCNY). For example, (0.9, 0.9): imputed quality-of-life values have been reduced by 10% in both arms. MAR: missing at random, MNAR: missing not at random, QALY: quality-adjusted life year, YCNY: Your Care Needs You.

4. Discussion

The study's findings provide detailed insights into the short-term costs and health outcomes associated with the YCNY intervention. In particular, the YCNY intervention was cost savings alongside slight improvements in health outcomes compared to usual care. Although the observed changes in costs and outcomes were minor, this analysis indicates that the YCNY intervention may offer a viable and efficient means of achieving health benefits in resource-constrained environments.

Recent economic studies conducted after Nuckols' systematic review have evaluated several transitional care programs (TCPs) across different healthcare systems. For example, studies in the US (Heo et al., 2023) and Canada (Murmann et al., 2023) have indicated that TCPs can reduce healthcare consumption and associated costs, with implications in the cost-effectiveness of these programs. However, these studies had limitations, such as not using health-related quality of life (HRQoL) as an outcome measure, being single-centre studies, and employing quasi-experimental designs. Nevertheless, one UK-based study by Sahora et al. (2017) presents similarities to our research. This study evaluated the cost-effectiveness of a Community In-reach Rehabilitation and Care Transition (CIRACT) service compared to traditional hospital-based rehabilitation (THB-Rehab) through a randomised controlled trial involving 250 participants. The trial measured hospital length of stay, unplanned readmission rates at day 28 and day 91, and HRQoL using the EQ-5D-3L at 91 days post-discharge. Their findings indicated an incremental cost of CIRACT over THB-Rehab of £144 (95% CI: -1645 to 1934) and an incremental QALY gain of 0.0404 (95% CI: -0.0566 to 0.1375), resulting in an incremental cost-effectiveness ratio (ICER) per QALY of £2,022 (95% CI: -76,895 to 121,856). However, the study had some limitations and differences with our study, including its single-centre nature (the Queen's Medical Centre, a large teaching hospital in Nottingham, UK), the relative youngest patient population (adults aged 70 years and older), the distant recruitment period (2013-2014), the small sample size, and the no clear statistical management of missing data.

Our study has several strengths. Firstly, our multicentre design, encompassing 15 hospitals and 35 wards, ensures that our results are more representative than those single-centre studies. Secondly, although our sample size might be considered modest, it is nearly double that of the Sahora et al. study, providing greater statistical power and confidence in our results. Thirdly, we employed current recommended methodologies to address common statistical challenges in economic evaluation using randomised trial designs (El Alili et al., 2022). For instance, many economic evaluations using data from cluster-randomised trials overlook the importance of accounting for clustering or correlation which can lead to underestimated statistical uncertainty and inaccurate point estimates (El Alili et al., 2020; Gomes et al., 2012b; Grieve et al., 2010). In our study, we addressed the hierarchical structure of the data and the potential correlation between costs and effects at the ward level. Additionally, we addressed a common issue in economic evaluations: the handling of missing data. Most studies discard observations with missing data and report only complete case analyses (CCAs), which can introduce bias if the missingness is related to unobserved values (missing not at random, or MNAR) (Fiero et al., 2016; Gabrio et al., 2017; Mutubuki et al., 2021; Noble et al., 2012). For instance, patients in poorer health may be less likely to complete quality-of-life questionnaires, skewing results. To mitigate this, we implemented a multiple imputation exercise using chained equations and conducted an extensive scenario analysis, providing a more comprehensive analysis.

Despite these strengths, our study has limitations that should be acknowledged. First, the partial availability of EQ-5D scores from a smaller subset of the patient cohort (collected via CRF from individuals) introduces uncertainty, compounded by substantial missing data on these measurements. Additionally, most of the healthcare resource consumption information (with the exception of hospitalisation) was collected via CRFs. These resources include certain hospital services, primary care services, and social services. Unfortunately, data on these non-hospitalisation resources exhibited a high percentage of missing values and were, therefore, excluded from the primary cost-effectiveness analysis. However, an analysis of healthcare consumption by trial arm suggests that the control group utilised more primary care and social services compared to the intervention group. Consequently, we believe our cost-effectiveness results may be considered conservative estimates. Third, in our scenario analysis addressing the missing HRQoL data, we rescaled the MAR-imputed HRQoL using a multiplicative factor. However, alternative pattern-mixture approaches, such as applying an additive factor, may have been considered to 'offset' the data. Moreover, with longitudinal data, the assumption of constant departure from MAR over time, as used in this study, may not hold. For instance, the departure could vary with time since withdrawal, potentially increasing over time (Ratitch et al., 2013)**.** Fourth, one challenge with conducting an MNAR scenario analysis is selecting plausible sensitivity parameter values. An alternative approach could be to formally elicit expert beliefs regarding the missing data distribution (White et al., 2007). These experts may include trial investigators, clinicians, or even patients who can provide valuable insights into the missing data mechanisms. Mason et al. (2017), for example, offer a useful framework for eliciting expert opinion on MNAR mechanisms in cost-effectiveness analysis (CEA). Finally, the 90-day follow-up period may not fully capture the medium-term impacts of the YCNY intervention. This limitation is particularly important, as medium- and long-term effects may significantly influence the cost-effectiveness outcomes. Therefore, a long-term economic evaluation, considering the lifetime horizon of these patients, is necessary to thoroughly assess the YCNY intervention.

In conclusion, the YCNY intervention represents a promising, cost-effective approach for improving the safety and experience of older adults transitioning from hospital to home in the UK. These findings support the integration of patient-centred care models like YCNY into healthcare policies, potentially fostering more sustainable and effective care strategies within the NHS. By adopting such interventions, the healthcare system can strive towards better patient outcomes and efficient resource utilisation. Continued research and investment in innovative transitional care programs are essential to realising these benefits.

**Disclaimer**

The views expressed in this publication are those of the author(s) and not necessarily those of the NHS, the National Institute for Health Research or the Department of Health and Social Care.

**Funding**

This summary presents independent research funded by the National Institute for Health Research (NIHR) under the NIHR Programme Grant for Applied Research (PGfAR) Award, reference RP-PG-1214-20017. Partners at Care Transitions (PACT): Improving Patient Experience and Safety at Transitions of Care.

**Availability of data and materials**

The datasets used and analysed during the current study are available from the corresponding author on reasonable request.

**Competing interests**

The authors declare that they have no competing interests.

**Supplementary material**

**Table S1**. CHEERS 2022 Checklist.

| Topic | No. | Item | Location where item is reported |
|---|---|---|---|
| **Title** | | | |
| Title | 1 | Identify the study as an economic evaluation and specify the interventions being compared. | Title, Page 1 |
| **Abstract** | | | |
| Abstract | 2 | Provide a structured summary that highlights context, key methods, results, and alternative analyses. | Abstract, Page 1 |
| **Introduction** | | | |
| Background and objectives | 3 | Give the context for the study, the study question, and its practical relevance for decision making in policy or practice. | Introduction, Paragraph 1-3 |
| **Methods** | | | |
| Health economic analysis plan | 4 | Indicate whether a health economic analysis plan was developed and where available. | Methods, Paragraph 2 |
| Study population | 5 | Describe characteristics of the study population (such as age range, demographics, socioeconomic, or clinical characteristics). | Methods, Paragraph 6; Results, Table 1 |
| Setting and location | 6 | Provide relevant contextual information that may influence findings. | Methods, Paragraph 3 |
| Comparators | 7 | Describe the interventions or strategies being compared and why chosen. | Methods, Paragraph 4 |
| Perspective | 8 | State the perspective(s) adopted by the study and why chosen. | Methods, Paragraph 5 |
| Time horizon | 9 | State the time horizon for the study and why appropriate. | Methods, Paragraph 5 |
| Discount rate | 10 | Report the discount rate(s) and reason chosen. | Methods, Paragraph 5 |
| Selection of outcomes | 11 | Describe what outcomes were used as the measure(s) of benefit(s) and harm(s). | Methods, Paragraph 6 |
| Measurement of outcomes | 12 | Describe how outcomes used to capture benefit(s) and harm(s) were measured. | Methods, Paragraph 6 |
| Valuation of outcomes | 13 | Describe the population and methods used to measure and value outcomes. | Methods, Paragraph 6 |
| Measurement and valuation of resources and costs | 14 | Describe how costs were valued. | Methods, Paragraphs 7–8 |
| Currency, price date, and conversion | 15 | Report the dates of the estimated resource quantities and unit costs, plus the currency and year of conversion. | Methods, Paragraph 7 |
| Rationale and description of model | 16 | If modelling is used, describe in detail and why used. Report if the model is publicly available and where it can be accessed. | Not applicable |
| Analytics and assumptions | 17 | Describe any methods for analysing or statistically transforming data, any extrapolation methods, and approaches for validating any model used. | Not applicable |
| Characterising heterogeneity | 18 | Describe any methods used for estimating how the results of the study vary for subgroups. | Not applicable |
| Characterising distributional effects | 19 | Describe how impacts are distributed across different individuals or adjustments made to reflect priority populations. | Not applicable |
| Characterising uncertainty | 20 | Describe methods to characterise any sources of uncertainty in the analysis. | Not applicable |
| Approach to engagement with patients and others affected by the study | 21 | Describe any approaches to engage patients or service recipients, the general public, communities, or stakeholders (such as clinicians or payers) in the design of the study. | Not applicable |
| **Results** | | | |
| Study parameters | 22 | Report all analytic inputs (such as values, ranges, references) including uncertainty or distributional assumptions. | Results, Paragraph 1; Tables 1–4 |
| Summary of main results | 23 | Report the mean values for the main categories of costs and outcomes of interest and summarise them in the most appropriate overall measure. | Results, Paragraphs 2–3 |
| Effect of uncertainty | 24 | Describe how uncertainty about analytic judgments, inputs, or projections affect findings. Report the effect of choice of discount rate and time horizon, if applicable. | Sensitivity Analysis, Table 5; Figure S2. |
| Effect of engagement with patients and others affected by the study | 25 | Report on any difference patient/service recipient, general public, community, or stakeholder involvement made to the approach or findings of the study | Not applicable |
| **Discussion** | | | |
| Study findings, limitations, generalisability, and current knowledge | 26 | Report key findings, limitations, ethical or equity considerations not captured, and how these could affect patients, policy, or practice. | Discussion, Paragraphs 1–3 |
| **Other relevant information** | | | |
| Source of funding | 27 | Describe how the study was funded and any role of the funder in the identification, design, conduct, and reporting of the analysis | Funding section |
| Conflicts of interest | 28 | Report authors conflicts of interest according to journal or International Committee of Medical Journal Editors requirements. | Competing interests section |

From: Husereau D, Drummond M, Augustovski F, et al. Consolidated Health Economic Evaluation Reporting Standards 2022 (CHEERS 2022) Explanation and Elaboration: A Report of the ISPOR CHEERS II Good Practices Task Force. Value Health 2022;25. doi:10.1016/j.jval.2021.10.008

**Table S2**. Sample sizes across datasets.

| Dataset | Initial sample | Primary analysis cohort sample | Primary analysis cohort sample + patients who have died within 30 days and no readmission recorded |
|---|---|---|---|
| Routine | 5,450 | 4,947 | 5,147 |
| CRF | 622 | 455 | 468 |
| EE | 615 | 455 | 468 |

Notes: CRF: Complete case form; EE: Economic evaluation.

**Table S3.** Intervention delivery per patient.

| Activity | Duration (minutes) | | | Undertaken by |
|---|---|---|---|---|
| | Usual care wards | Intervention wards | Incremental | |
| Discussion with patient about care - on admission | 5 | 5 | 0 | Nursing/Manager |
| Discussion with patient about care - during admission | 15 | 30 | 15 | Nursing/Manager |
| Discussion with patient about care - during admission | 15 | 25 | 10 | Medical staff |
| Discussion with patient about care - discharge | 15 | 20 | 5 | Nursing/Manager |
| Assisting with activities of daily living | 70 | 105 | 35 | Nursing/Manager |
| Instructions/education for patient and/or caregiver | 5 | 25 | 20 | Nursing/Manager |

**Table S4**. Comparison between the EE dataset and the Routine dataset at baseline.

| Variable | Routine dataset | EE dataset |
|---|---|---|
| | Patients=5,147<br>Wards=39 | Patients=468<br>Wards=35 |
| Age | 84.5 (5.8) | 83.1 (5.4) |
| Dummy sex, 1=Male, % | 41.4 (49.3) | 43.6 (49.6) |
| Ward: baseline readmission, % | 18.0 (5.3) | 19.1 (6.2) |
| Ward: patients over 75 yo, % | 74.7 (22.5) | 72.0 (23.7) |
| Ward: Dummy specialty, 1=Eldery & interm. care, % | 59.0 (49.2) | 54.5 (49.9) |

**Table S5**. Missing values.

| Variable | Missing (n) | Total | Missing (%) |
|---|---|---|---|
| *Baseline data* | | | |
| Intervention | 0 | 468 | 0 |
| Age | 0 | 468 | 0 |
| Sex | 0 | 468 | 0 |
| EQ-5D at baseline | 9 | 468 | 1.92 |
| Ward specialty | 0 | 468 | 0 |
| Baseline readmission rate | 12 | 468 | 2.56 |
| Patients over 75 years old | 0 | 468 | 0 |
| *Quality of life* | | | |
| EQ-5D at 10 days | 149 | 468 | 31.84 |
| EQ-5D at 30 days | 159 | 468 | 33.97 |
| EQ-5D at 90 days | 213 | 468 | 45.51 |
| *Resource use - Hospital care* | | | |
| Hospitalisation | 2 | 468 | 0.43 |
| Outpatient clinic | 284 | 468 | 60.68 |
| Day case | 367 | 468 | 78.42 |
| A&E | 359 | 468 | 76.71 |
| *Resource use - Primary care* | | | |
| GP at surgery | 289 | 468 | 61.75 |
| GP at home | 313 | 468 | 66.88 |
| GP at telephone | 295 | 468 | 63.03 |
| Nurse at surgery | 298 | 468 | 63.68 |

| Nurse at home | 302 | 468 | 64.53 |
| Nurse at telephone | 318 | 468 | 67.95 |
| Therapist | 300 | 468 | 64.10 |
| *Resource use - Social care* | | | |
| Home care | 304 | 468 | 64.96 |
| Social worker | 308 | 468 | 65.81 |

**Table S6.** Unadjusted mean intervention, healthcare and total costs between arms. Monetary values expressed in British pounds of 2022. Results based on imputed values.

| Variable | Intervention | Control | Difference |
|---|---|---|---|
| Intervention costs | 94.32 | 0.00 | 94.32 |
| Hospitalisation costs | 1,502.44 | 1,812.44 | -310.00 |
| Total costs | 1,596.76 | 1,812.44 | -215.68 |

**Table S7**. Other healthcare resource use within 90 days by arm and total, complete case analysis.

| Healthcare resource use | Intervention | | Control | | Total | |
|---|---|---|---|---|---|---|
| | N | % | N | % | N | % |
| *Hospital care* | | | | | | |
| Outpatient clinic | 84 | 66.7 | 100 | 68 | 184 | 67.4 |
| Day case | 42 | 14.3 | 59 | 23.7 | 101 | 19.8 |
| A&E | 45 | 20 | 64 | 29.7 | 109 | 25.7 |
| *Primary care* | | | | | | |
| GP at surgery | 82 | 51.2 | 97 | 54.6 | 179 | 53.1 |
| GP at home | 72 | 22.2 | 83 | 31.3 | 155 | 27.1 |
| GP at telephone | 83 | 62.6 | 90 | 70 | 173 | 66.5 |
| Nurse at surgery | 79 | 43 | 91 | 49.4 | 170 | 46.5 |
| Nurse at home | 80 | 52.5 | 86 | 54.6 | 166 | 53.6 |
| Nurse at telephone | 74 | 33.8 | 76 | 27.6 | 150 | 30.7 |
| Therapist | 81 | 48.1 | 87 | 44.8 | 168 | 46.4 |
| *Social care* | | | | | | |
| Home care | 79 | 40.5 | 85 | 51.8 | 164 | 46.3 |
| Social worker | 81 | 17.3 | 79 | 24 | 160 | 20.6 |

**Table S8**. Missing data pattern for the EQ-5D-3L and QALYs variables by intervention and control arms.

| Arm: Intervention | | | |
|---|---|---|---|
| Variable | Missing | Total | % missing |
| EQ-5D at Baseline | 0 | 222 | 0 |
| EQ-5D at 10 days | 59 | 222 | 26.58 |
| EQ-5D at 30 days | 65 | 222 | 29.28 |
| EQ-5D at 90 days | 82 | 222 | 36.94 |
| Arm: Control | | | |
| Variable | Missing | Total | % missing |
| EQ-5D at Baseline | 0 | 246 | 0 |
| EQ-5D at 10 days | 59 | 246 | 23.98 |
| EQ-5D at 30 days | 73 | 246 | 29.67 |
| EQ-5D at 90 days | 94 | 246 | 38.21 |

**Figure S1.** Pattern of missing data for the EQ-5D-3L and QALYs variables.

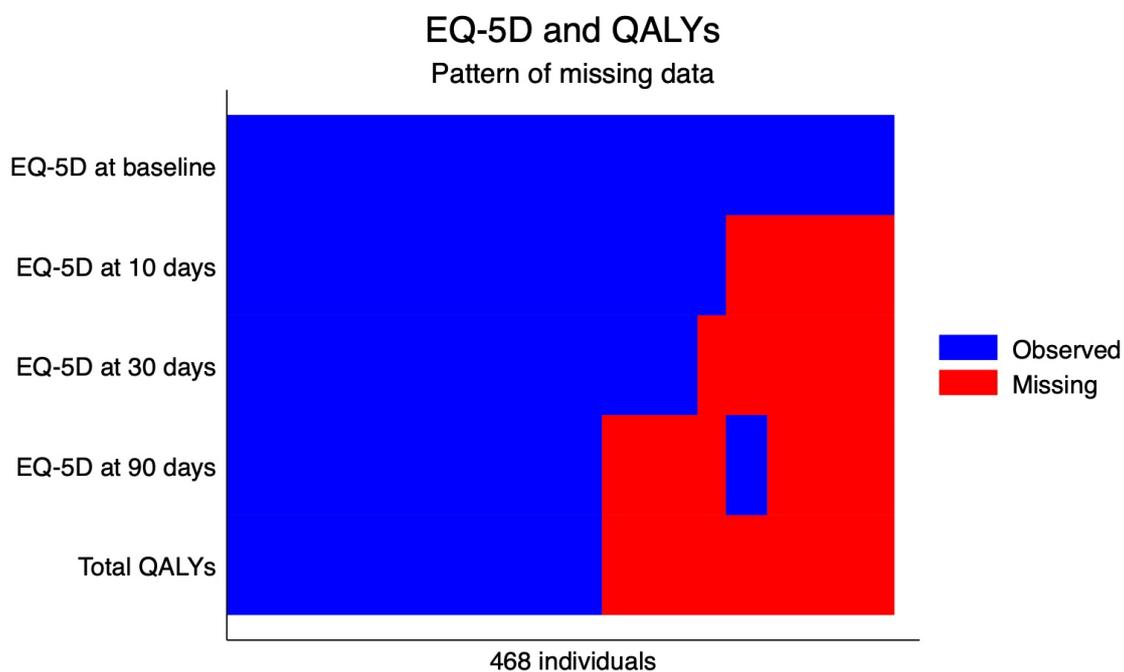

**Table S9.** Multilevel mixed-effect logistic estimations for the missingness in EQ-5D-3L measurements.

| Variable | Dummy missing EQ-5D at 10 days, 1=Yes | Dummy missing EQ-5D at 30 days, 1=Yes | Dummy missing EQ-5D at 90 days, 1=Yes |
|---|---|---|---|
| Dummy intervention arm | 0.161 | 0.0544 | -0.136 |
|  | (0.74) | (0.27) | (-0.76) |
| EQ-5D at baseline | -1.018** | -0.986*** | -0.0126 |
|  | (-3.20) | (-3.44) | (-0.04) |
| Dummy ward speciality, 1=Eldery and intermediate care | 0.471 | 0.278 | 0.134 |
|  | (1.12) | (0.66) | (0.33) |
| Ward: baseline readmission, % | 0.00225 | -0.00128 | -0.00875 |
|  | (0.15) | (-0.11) | (-0.60) |
| Ward: patients over 75 yo, % | -0.00686 | 0.000618 | 0.000955 |
|  | (-1.17) | (0.10) | (0.12) |
| Dummy sex, 1=Male | -0.129 | -0.259 | 0.102 |
|  | (-0.58) | (-1.52) | (0.54) |
| Constant | -0.418 | -0.471 | -0.437 |
|  | (-0.89) | (-1.31) | (-1.02) |
| N | 468 | 468 | 468 |

Notes: t statistics in parentheses. *p<0.05, **p<0.01, ***p<0.001.

**Table S10.** Unadjusted mean total costs and total QALYs by arm.

|  |  | Intervention | Control |
|---|---|---|---|
| Total cost | Mean | 1,636.77 | 1,869.58 |
|  | 95% CI | 1,210.40 - 2,063.13 | 1,444.54 - 2,294.62 |
| Total QALYs | Mean | 0.1459 | 0.1354 |
|  | 95% CI | 0.1369 - 0.1548 | 0.1269 - 0.1439 |

**Figure S2.** Cost-effectiveness acceptability curves.

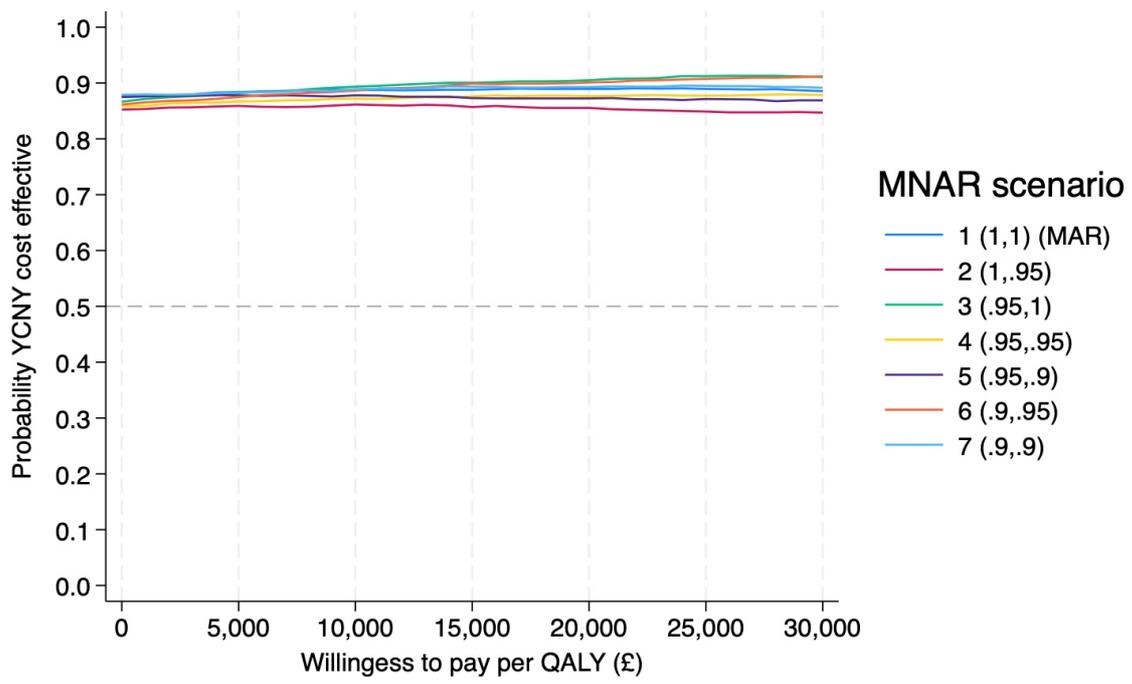